\documentclass[10pt, conference]{IEEEtran}

\usepackage{amsmath,amssymb,amsfonts}
\usepackage{algorithm}
\usepackage[noend]{algpseudocode}
\usepackage{cite}
\usepackage{graphicx}
\usepackage{textcomp}
\usepackage{subcaption}
\usepackage[svgnames]{xcolor}
\usepackage{multirow}
\usepackage{hyperref}
\usepackage{makecell}
\usepackage{booktabs}
\usepackage{listings}
\usepackage{tikz}
\usepackage{pifont}
\usepackage{comment}
\usepackage{enumitem}
\usepackage{scalerel}
\usepackage{makecell}
\usepackage{caption}
\usepackage{float}
\usepackage[outline]{contour}
\usepackage{bm}
\usepackage{utfsym}
\usepackage{setspace}
\usepackage[utf8x]{inputenc}

\hypersetup{colorlinks,
    linkcolor=ACMRed,
    citecolor=ACMPurple,
    urlcolor=ACMDarkBlue,
    filecolor=ACMDarkBlue}

\definecolor[named]{ACMBlue}{cmyk}{1,0.1,0,0.1}
\definecolor[named]{ACMYellow}{cmyk}{0,0.16,1,0}
\definecolor[named]{ACMOrange}{cmyk}{0,0.42,1,0.01}
\definecolor[named]{ACMRed}{cmyk}{0,0.90,0.86,0}
\definecolor[named]{ACMLightBlue}{cmyk}{0.49,0.01,0,0}
\definecolor[named]{ACMGreen}{cmyk}{0.20,0,1,0.19}
\definecolor[named]{ACMPurple}{cmyk}{0.55,1,0,0.15}
\definecolor[named]{ACMDarkBlue}{cmyk}{1,0.58,0,0.21}
\definecolor[named]{baselinemax}{rgb}{1,0.647058824,0}

\definecolor{codegreen}{rgb}{0,0.6,0}
\definecolor{codegray}{rgb}{0.5,0.5,0.5}
\definecolor{codepurple}{rgb}{0.58,0,0.82}
\definecolor{codebgcolor}{rgb}{0.95,0.95,0.92}
\lstdefinestyle{codestyle}{
    backgroundcolor=\color{codebgcolor},
    commentstyle=\color{codegreen},
    keywordstyle=\color{magenta},
    numberstyle=\tiny\color{codegray},
    stringstyle=\color{codepurple},
    basicstyle=\linespread{0.80}\ttfamily\footnotesize,
    breakatwhitespace=false,         
    breaklines=true,                 
    captionpos=b,                    
    keepspaces=true,                 
    showspaces=false,                
    showstringspaces=false,
    showtabs=false,                  
    tabsize=2,
    escapechar=\%,
    columns=fullflexible,
    framexleftmargin=4pt,
}
\lstnewenvironment{cpp}{\lstset{style=codestyle,language={C++}}}{}

\usepackage[all=normal,paragraphs=tight,floats=tight,mathspacing=tight,wordspacing=tight,leading=tight]{savetrees}
\usepackage[stretch=0,shrink=100]{microtype}
\newcommand{\ls}{{LightningSim}}

\newcommand{\ourwork}{{FIFOAdvisor}}

\title{\huge\ourwork: A DSE Framework for Automated\\FIFO Sizing of High-Level Synthesis Designs\vspace{-0.25em}}

\author{
\IEEEauthorblockN{Stefan Abi-Karam\IEEEauthorrefmark{1}\IEEEauthorrefmark{2}, 
Rishov Sarkar\IEEEauthorrefmark{1}, 
Suhail Basalama\IEEEauthorrefmark{3}, 
Jason Cong\IEEEauthorrefmark{3}, 
Callie Hao\IEEEauthorrefmark{1}}

\IEEEauthorblockA{\IEEEauthorrefmark{1}Georgia Institute of Technology, \IEEEauthorrefmark{2}Georgia Tech Research Institute,\IEEEauthorrefmark{3}University of California, Los Angeles}

\IEEEauthorblockA{\{stefanabikaram, rishov.sarkar, callie.hao\}@gatech.edu, \{basalama, cong\}@cs.ucla.edu}\vspace{-2em}
}

\begin{document}

\maketitle

\begin{abstract}

Dataflow hardware designs are important for efficient algorithm implementations on FPGAs across various domains using high-level synthesis (HLS). However, these designs pose a challenge: correctly and optimally sizing first-in-first-out (FIFO) channel buffers. FIFO sizes are user-defined parameters, introducing a trade-off between latency and area—undersized FIFOs cause stalls and increase latency, while oversized FIFOs waste on-chip memory. In many cases, insufficient FIFO sizes can also lead to deadlocks.
Deciding the best FIFO sizes is non-trivial. Existing methods make limiting assumptions about FIFO access patterns, overallocate FIFOs conservatively, or use time-consuming RTL simulations to evaluate different FIFO sizes. Furthermore, we highlight that runtime-based analyses (i.e., simulation) are the only way to solve the FIFO optimization problem while ensuring a deadlock-free solution for designs with data-dependent control flow.

To tackle this challenge, we propose \ourwork{}, a framework to automatically decide FIFO sizes in HLS designs. Our approach is powered by LightningSim, a fast simulator that is 99.9\% cycle-accurate and supports millisecond-scale incremental simulations with new FIFO configurations. We formulate FIFO sizing as a dual-objective black-box optimization problem and explore various heuristic and search-based methods to analyze the latency–resource trade-off. We also integrate \ourwork{} with Stream-HLS, a recent framework for optimizing affine dataflow designs lowered from C++, MLIR, or PyTorch, enabling deeper optimization of the heavily-used FIFOs in these workloads.

We evaluate \ourwork{} on a suite of Stream-HLS benchmarks, including linear algebra and deep learning workloads, to demonstrate our approach's ability to optimize large and dynamic dataflow patterns. Our results show Pareto-optimal latency–memory usage frontiers for FIFO configurations generated via different optimization strategies. Compared to baseline designs with naïvely-sized FIFOs, \ourwork{} identifies configurations with much lower memory usage and minimal delay overhead. Additionally, we measure the runtime of our optimization process and demonstrate significant speedups compared to traditional HLS/RTL co-simulation-based approaches, making \ourwork{} practical for rapid design space exploration. Finally, we present a case study using \ourwork{} to optimize a complex hardware accelerator with non-trivial data-dependent control flow.

Code and results open-sourced at \url{https://github.com/sharc-lab/fifo-advisor}.
\end{abstract}

\section{Introduction}

Dataflow hardware designs play an important role in domain-specific accelerators~\cite{tapa,abi2023inr,lerner2024data,basalama2025stream,nowatzki2017stream,kim2016dataflow,sarkar2023flowgnn} including systolic arrays \cite{wei2017automated, basalama2023comprehensive, basalama2022versatile, wang2021search, polysa}, such as AutoSA \cite{autosa} and SuSy \cite{susy}. They allow different tasks to execute simultaneously while communicating through first-in, first-out (FIFO) channels and thus improve overall design performance such as latency and throughput. 

In designing dataflow architectures using high-level synthesis (HLS), users typically define parallel tasks as functions, which are synthesized into hardware modules, and use FIFO streams to pass data between them. One important consideration is to correctly and optimally decide the depth of each FIFO stream, which has a significant impact on the overall design quality.
Small FIFO sizes may result in deadlocks or severe stalls that largely hurt design performance, while large FIFO sizes require large on-chip memory capacity with possible resource and area waste.

Determining the optimal sizes of FIFOs is very challenging. First, the design space can be exponentially large with the growing number of FIFOs in the design, especially given the intertwined effects of multiple FIFOs that cannot be statically analyzed. For example, increasing the sizes of one or more FIFOs does not necessarily improve the design performance.
Second, exploring the trade-off between latency and memory is nonlinear and difficult to model. Without proper simulation and synthesis, it is impossible to accurately estimate the performance and memory change because of FIFO size change.

Prior works addresses the FIFO sizing problem by enforcing producer–consumer rate matching by design, as in systolic array architectures (e.g., PolySA~\cite{polysa} and AutoSA~\cite{autosa}), which assume constant FIFO throughput~\cite{cong_2014}. Meanwhile, other buffer sizing strategies remain incompatible with traditional HLS design flows (e.g., dynamically scheduled HLS~\cite{dyn_hls}). More fundamentally, static-analysis-based methods are inherently restricted to designs that \textit{do not} include data-dependent control flow, a characteristic common in many modern HLS designs, which limits the overall applicability~\cite{ha2019decidable} of prior approaches.

\begin{figure*}
    \centering
    \includegraphics[width=0.95\linewidth]{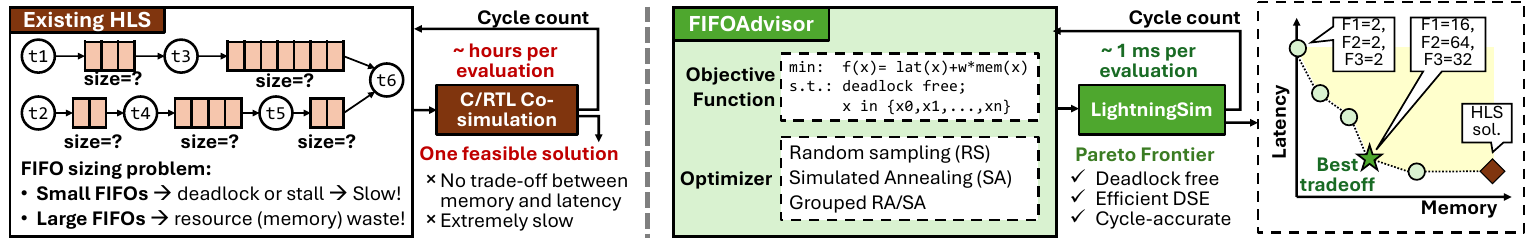}
    \vspace{-0.1em}
    \caption{Overview of \textbf{\ourwork{}}, highlighting the limitations of previous FIFO sizing techniques and the advantages of our multi-objective optimization approach using \ls{} for fast, cycle-accurate simulation. \ourwork{} generates a Pareto frontier of candidate solutions that trade off between memory usage reduction and latency overhead.}
    \label{fig:overview}
\end{figure*}

Instead, existing tools and other prior work usually sidestep this problem either by overallocating FIFOs or using time-consuming RTL simulations to test different FIFO sizes. For instance, Stream-HLS~\cite{streamhls} is a state-of-the-art work that automatically generates HLS dataflow designs for machine learning kernels, but it maximizes FIFO sizes to their respective write counts because of the complexity of choosing these sizes correctly. Meanwhile, Vitis HLS has a tool to repeatedly run RTL simulation for an HLS design with higher and higher FIFO sizes until it no longer deadlocks. As shown in Fig.~\ref{fig:overview} left, this makes it very challenging to find one feasible solution, let alone explore the latency/memory trade-off---a much harder problem, given the entangled effects of multiple FIFOs.

Addressing the complexity of the problem and limitations of existing tools, we propose \textbf{\ourwork{}}, a novel framework for design space exploration (DSE) of FIFO sizes for HLS dataflow designs, depicted in Fig.~\ref{fig:overview}. Our work features:

\setlist[itemize]{leftmargin=1em}
\begin{itemize}
    \item \textbf{Rapid Automated DSE:} \ourwork{} is the first work that can find a Pareto frontier of FIFO sizes for arbitrary HLS dataflow designs quickly with push-button ease of use---often in under one minute end-to-end. This is powered by LightningSim~\cite{lightningsim}, which performs incremental simulations in under 1\,ms on average per FIFO size change.

     \item \textbf{Comprehensive Evaluation with Multiple Search Algorithms:} We introduce several optimization methods to guide DSE, analyzing each optimizer’s efficiency and solution quality for latency/memory trade-offs. Our framework achieves search runtimes of seconds, versus hours or days for traditional HLS/RTL co-simulation. We also present a case study using \ourwork{} to optimize a non-trivial hardware accelerator with data-dependent control flow.
    
    \item \textbf{Open-Source and Integration with Stream-HLS:}
    We integrate \ourwork{} with Stream-HLS~\cite{streamhls} to automatically optimize the FIFOs in the generated dataflow HLS kernels, and we evaluate \ourwork{} on 22 of these kernels, representative of real machine learning applications. Our tool can also be used in a standalone fashion, and we open-source all our code and results.
\end{itemize}

\section{Motivation \& Prior Work}
\label{sec:motivation-prior-work}

\subsection{Static vs.\ Runtime Analysis}

To solve the deadlock-free FIFO sizing optimization problem for a broad class of real-world HLS designs, specifically those exhibiting data-dependent control flow, static analysis alone is insufficient~\cite{ha2019decidable}. Static methods cannot model the control flow that might conditionally enable FIFO read and write operations. Therefore, we argue that runtime analysis methods, such as the simulation-based approach proposed in this work, are the only way to find deadlock-free solutions for designs with data-dependent control flow. This is in contrast to prior efforts in FIFO and buffer sizing for HLS as we note in Table~\ref{tab:prior-work-comparison}.

As a supporting example, consider the design in Figure \ref{fig:example-hls-design}. The only way to choose optimal depths for FIFOs \texttt{x} and \texttt{y} without possibly incurring a deadlock is to know the value of input argument, \texttt{n}, passed into the kernel at runtime, information that is not present during static analysis. This example motivates a larger family of common HLS design patterns found in real-world HLS accelerator designs in which data-dependent control flow interacts with FIFO operations, and where only a dynamic, simulation-driven approach can guarantee correct, minimal FIFO sizing.

\subsection{Prior Work}

The problem of optimizing buffer and FIFO sizes in HLS has previously been addressed, but with major limitations that prevent their application to a wide range of practical ``real world'' HLS designs.

\textbf{Static Analysis with Synchronous Dataflow (SDF) Models.} 
Existing work on SDF buffer sizing relies on static analysis combined with a system of difference constraints (SDC) optimization model to determine optimal FIFO sizes for HLS designs~\cite{cong_2014}. However, this approach assumes  FIFOs with fixed read and write rates, resulting in constant throughput between dataflow processes. SDF-based analysis methods cannot handle irregular FIFO read and write patterns or data-dependent control flow, which are ubiquitous in modern HLS designs. For example, Stream-HLS designs exhibit highly irregular FIFO read and write timing patterns that violate the design assumptions required by SDF analysis.

\textbf{Dynamic HLS Buffer Optimization.} 
Recent dynamic HLS work uses static analysis, dataflow rewrites, and MILP to optimize buffer sizes~\cite{dyn_hls}. While this appears to address the limitations of SDF analysis, it targets a fundamentally different problem. In dynamic HLS, buffers are implicit in a token-based dataflow protocol: reading from an empty buffer is a no-op, and only writes to full buffers cause stalls. In contrast, Vitis HLS streams have explicit, potentially blocking reads and writes, with timing controlled by data-dependent control flow. Thus, buffer sizing methods for dynamic HLS are not applicable to deadlock-free explicit FIFO sizing in traditional HLS designs, making these optimization techniques inapplicable to the scope of this work.

\textbf{Need for Runtime Analysis.}
Even if we ignore the shortcomings of previous methods and try to apply SDC or MILP-based optimization, these techniques still fail because they require static analysis of the HLS program to formulate the optimization problem. With data-dependent or irregular control flow, static analysis cannot capture FIFO access patterns, making such optimization approaches infeasible.

Given the limitations of static analysis and prior work, simulation-based runtime analysis is essential for optimizing FIFO sizes without deadlocks in designs with data-dependent control flow. This motivates our use of LightningSim \cite{lightningsim} to enable rapid evaluation of candidate FIFO configurations with fast simulation and runtime program analysis, making it the only practical solution for a large class of HLS designs.

\begin{table}[t!]
\renewcommand{\arraystretch}{1.1}
\setlength{\tabcolsep}{4pt}
\centering
\footnotesize

\resizebox{0.90\columnwidth}{!}{
\begin{tabular}{l|lcccc}
\toprule
\textbf{Prior Work} &
\textbf{Method} &
\textbf{CT} &
\textbf{IR/W} &
\textbf{DDCF} \\
\midrule
\textbf{Josipovic et al.~\cite{dyn_hls}} & \makecell[l]{Static Analysis + \\ Dataflow Rewrites + \\ MILP} & N/A& N/A & N/A \\
\hline
\textbf{Cong et al.~\cite{cong_2014}} & \makecell[l]{Static Analysis + \\ SDC} & {\normalsize\textcolor{Green}{\protect\usym{2713}}} & {\normalsize\textcolor{red}{\protect\usym{2715}}} & {\normalsize\textcolor{red}{\protect\usym{2715}}} \\
\hline
\textbf{\ourwork} & \makecell[l]{Simulation + \\ Black-Box Opt.} & {\normalsize\textcolor{Green}{\protect\usym{2713}}}& {\normalsize\textcolor{Green}{\protect\usym{2713}}} & {\normalsize\textcolor{Green}{\protect\usym{2713}}} \\
\bottomrule
\end{tabular}
}
\caption{Comparison of Prior Works Related to Buffer and FIFO Sizing. \textbf{CT} indicates support for optimizing FIFOs with constant throughput. \textbf{IR/W} indicates support for irregular FIFO read/write patterns. \textbf{DDCF} indicates support for data-dependent control flow.}
\label{tab:prior-work-comparison}
\end{table}

\begin{figure}
    \centering
    \begin{cpp}
#include <hls_stream.h>
using stream = hls::stream<int>;
void producer(stream &x, stream &y, int n) {
    for (int i = 0; i < n; i++) x.write(1);
    for (int i = 0; i < n; i++) y.write(1);
}
void consumer(int *out, stream &x, stream &y, int n) {
    int sum = 0;
    for (int i = 0; i < n; i++)
        sum += x.read() + y.read();
    *out = sum;
}
void mult_by_2(int *out, int n) {
    #pragma HLS dataflow
    stream x, y;
    producer(x, y, n);
    consumer(out, x, y, n);
}
\end{cpp}
   \vspace{-0.5em}
    \caption{HLS designs with FIFOs that can not be sized optimally or in a deadlock-free manner without runtime analysis (e.g. simulation)}
    \label{fig:example-hls-design}
\end{figure}

\section{Methodology}

The key challenge in optimizing FIFO sizes lies in the complex interdependencies that make it difficult to predict how changing one FIFO will impact the latency of downstream tasks and the overall design. Consider the dataflow design depicted in Fig.~\ref{fig:overview} left. Imagine that task \texttt{t1} is stalling frequently due to a full output FIFO. Will increasing that FIFO's depth improve the design's latency?
It depends on other tasks in the design---for instance, maybe the true bottleneck is \texttt{t5}'s output FIFO.

These inter-task effects are difficult to determine analytically, 
and doing so may even be impossible if the design has data-dependent control flow~\cite{ha2019decidable}. Instead, we formulate the FIFO sizing problem as a black-box multi-objective optimization task. Fig.~\ref{fig:overview} depicts our optimization flow.

Each candidate solution \( \mathbf{x} \) is a \textit{FIFO configuration}—a set of positive integer FIFO sizes, with a lower bound of $2$\footnote{A FIFO size of 2 is the smallest practical FIFO size, as a size of 1 stalls after the first write—likely the reason Vitis HLS defaults to 2, which we also use.}. Upper bounds, noted as $u$, are either the sizes defined in the design, the total number of writes observed during kernel execution, or user-specified.

Given \( \mathbf{x} \), \ourwork{} evaluates two objectives: kernel latency (in cycles), $f_{lat}$, using \ls{}, to be described in \S\ref{sec:lightningsim}, and FIFO BRAM usage, $f_{bram}$, using our model defined in \S\ref{sec:bram-model}, both as positive integers. These ``black-box'' functions are inexpensive to evaluate: each run takes milliseconds in single-evaluation mode, and under 1\,ms amortized in parallel mode.

We formalize the problem as:
\begin{gather*}
\label{eq:opt_obj}
\begin{array}{ll}
\text{Minimize} & \left( f_{\text{lat}}(\mathbf{x}),\ f_{\text{bram}}(\mathbf{x}) \right) \in \left(\mathbb{Z}_{\ge0},\ \mathbb{Z}_{\ge0}\right) \\[1pt]
& \operatorname{LightningSim}(\mathbf{x}) = \left( f_{\text{lat}}(\mathbf{x}),\ f_{\text{bram}}(\mathbf{x}) \right) \\[4pt]
\text{Subject to} & \text{No Deadlock}\\[1pt]
& \mathbf{x} = [x_1, x_2, \dots, x_n] \in \mathbb{Z}^n \quad \text{(FIFO sizes)} \\[1pt]
& 2 \leq x_i \leq u_i, \quad \forall i = 1, \dots, n 
\end{array}
\end{gather*}

We then have the freedom to build any optimizer to evaluate a set of FIFO configurations and minimize the dual objective. Pareto points are then selected from the optimizer-evaluated configurations.

To solve this optimization problem, we integrate LightningSim into the search procedure to obtain $f_{lat}$ (\S\ref{sec:lightningsim}), formulate $f_{bram}$ (\S\ref{sec:bram-model}), prune the search space whenever possible (\S\ref{sec:dse-bram-reduction}), and adapt classic optimizers to our problem (\S\ref{sec:optimizers}).

\subsection{LightningSim and Incremental Simulation}
\label{sec:lightningsim}

As previously mentioned, hardware simulators like Vitis HLS's built-in C/RTL co-simulation are far too slow for \ourwork{}: each simulation can take several minutes or even hours for large designs, meaning that evaluating a large number of design points would easily take hours or days.

Instead, we propose to utilize the state-of-the-art HLS simulator LightningSim~\cite{lightningsim}. LightningSim works by collecting an execution trace from the software execution of the HLS code and separately calculating its latency afterwards. This has the unique advantage that 
we can re-use the same execution trace to evaluate different FIFO sizes
---usually in less than \textbf{1\,ms} for the designs tested in this work outlined in Table \ref{tab:cosim}.

Thus, at the first invocation, LightningSim collects a full execution trace of a user's HLS design. We can then incrementally re-simulate the same design under any combination of FIFO sizes and obtain the changed latency $f_\text{lat}$ in under 1\,ms on average.

\subsection{Modeling Memory Usage}
\label{sec:bram-model}

While LightningSim provides an accurate model for $f_\text{lat}$, we also need a model for the other major variable impacted by changing FIFO sizes, namely, $f_\text{bram}$. We use FPGA BRAM utilization under Vitis HLS as our case study; extending this analysis to ASIC memory is trivial.

FIFOs with a depth of 2 or total size under 1K are implemented by shift registers and do not contribute to BRAM usage. Since BRAM resources are more scarce, we choose to primarily optimize for that metric while ignoring flip-flop overhead. Optimizing both BRAM and FF usage is in the scope of future work. For simplicity, we constrain our modeling to enforce that all FIFOs map exclusively to BRAMs, without inital support for URAM mapping. While extending our approach to URAMs is straightforward, we leave this as future work, with the same BRAM modeling methods directly applying to URAM modeling.

For other FIFOs, prior works~\cite{zhao_comba_2017,honorat2024automated} propose techniques to calculate the BRAM allocation. However, validated against exhaustive HLS synthesis runs, those techniques overestimate the true number of BRAMs allocated. Instead, our more accurate calculation, validated against these synthesis runs, is shown in Algorithm~\ref{alg:bram-algo}.

BRAM\_18K units support several configurations of row count and bitwidth: 1K\texttimes18, 2K\texttimes9, 4K\texttimes4, 8K\texttimes2, and 16K\texttimes1. We first allocate 1K\texttimes18 BRAMs for bitwidths that are multiples of 18 or higher (or sizes under 1K), then 2K\texttimes9 BRAMs satisfy the remaining multiples of 9 (or sizes under 2K), and so on.

Our model targets the UltraScale+ family of Xilinx FPGAs, but should generalize to other device families that use the BRAM18K primitive. We also have test harnesses to quickly evaluate Vitis HLS’s BRAM mapping behavior and can easily derive corresponding models for other device families, tool versions, or memory primitives (e.g., URAM).

\begin{algorithm}[t]
    \small
    \caption{Compute BRAM count for FIFO}
    \label{alg:bram-algo}
    \setstretch{0.8}
    \begin{algorithmic}
        \Require{FIFO depth $d$ and FIFO bitwidth $w$}
        \Ensure{BRAM count $n$}
        \State $n \gets 0$
        \If{$d \leq 2\ \vee\ dw \leq 1024$}
            \Return
        \EndIf
        \For{each supported BRAM depth $d_i$ and bitwidth $w_i$}
        \State $n \gets n + \left\lfloor{w/w_i}\right\rfloor\cdot\left\lceil{d/d_i}\right\rceil$ and $w \gets w\ \mathrm{mod}\ w_i$
        \If{$w > 0\ \wedge\ d \leq d_i$}
            \State $n \gets n + 1$ and $w \gets 0$
        \EndIf
        \EndFor
    \end{algorithmic}
\end{algorithm}

\subsection{Pruning the Design Space}
\label{sec:dse-bram-reduction}

We notice that this function $f_\text{bram}$ has a property that can drastically reduce the size of our search space. As noted in prior work~\cite{honorat2024automated}, since the BRAM usage increases in discrete steps at certain hard boundaries in FIFO depth, many FIFO sizes need not be considered. For instance, decreasing a FIFO's depth from 2048 to 2047 will never change the number of BRAMs but could potentially negatively impact the latency of the HLS design; therefore, we can skip testing the FIFO with depth 2047. We compute these breakpoints for each FIFO in an HLS design and limit our DSE to only those FIFO sizes that maximally utilize their allocated BRAMs.

\subsection{Included Optimizers}
\label{sec:optimizers}

With a fast simulator for latency and a model for BRAM usage, \ourwork{} proposes the following optimizers to solve the black-box multi-objective FIFO sizing problem.

\textbf{Random Sampling}: The simplest optimizer in \ourwork{} randomly samples a user-defined number of FIFO configurations, evaluates each, and extracts the Pareto-optimal points. Sampling random sizes in the range $2 \leq x_i \leq u_i$ is often ineffective, as only certain threshold sizes affect BRAM usage, as shown in \S\ref{sec:dse-bram-reduction}. Instead, we use our BRAM usage model to suggest optimal sizes for each FIFO, from which the sampler uniformly selects.

\textbf{Grouped Random Sampling}: A common HLS pattern is to define FIFO arrays, e.g., \texttt{hls::stream<float> data[16]}, suggesting parallel operations across FIFOs with similar access patterns and scheduling. Stream-HLS adopts this computation style, so we modify the random sampling optimizer to select a single FIFO depth for such groups rather than for each FIFO individually. As before, we use the BRAM usage model to suggest candidate sizes---identical to those suggested for individual FIFOs---from which one is randomly selected and applied to the entire group.

\textbf{Simulated Annealing}: When the design space is too large for individual or grouped random sampling (even after pruning) or has many infeasible designs, a guided search is needed to better evolve the Pareto frontier. To this end, \ourwork{} employs a simulated annealing optimizer that, like the random approaches, selects an index for each FIFO corresponding to one of the sizes from the pruned space.

To address the multi-objective nature of the search, we define a new single objective function as a weighted sum:
\allowdisplaybreaks
\begin{gather*}
f(\mathbf{x}) = (1-\beta)f_{\text{lat}}(\mathbf{x}) + \beta f_{\text{bram}}(\mathbf{x})\\
\text{for}\ \beta=\{0, 1/N, 2/N, ..., 1\}
\end{gather*}

The user specifies $N$, the number of $\beta$ values to search, which are linearly interpolated between 0 and 1. The optimizer launches $N$ simulated annealing runs, one for each $\beta$, optimizing the corresponding weighted objective function. After all runs are completed, all the evaluated points across runs are aggregated, and the Pareto frontier is extracted.

\textbf{Grouped Simulated Annealing}: Analogous to grouped random sampling, we extend the simulated annealing optimizer to operate on groups of FIFOs defined as arrays, assigning a shared FIFO depth to the entire group.

\textbf{Greedy Search}: FIFOs are often na\"ively sized to match a large input or output buffer, or to connect faster upstream computation to slower downstream computations—cases where a minimal depth of 2 often incurs no latency penalty. In such scenarios, a simple heuristic can efficiently optimize FIFO sizes. INR-Arch~\cite{inrarch} proposes iteratively reducing each FIFO’s depth to 2, ranked from largest to smallest observed depth during simulation, and simulating to check for deadlocks or excessive latency (defined as a fixed percentage over baseline). If a deadlock is triggered or an unreasonable increase in latency occurs, the original depth is restored. \ourwork{} adopts this as the greedy search optimizer.

\section{Evaluations}
\begin{figure*}[t]
    \centering
    \includegraphics[width=0.30\linewidth]{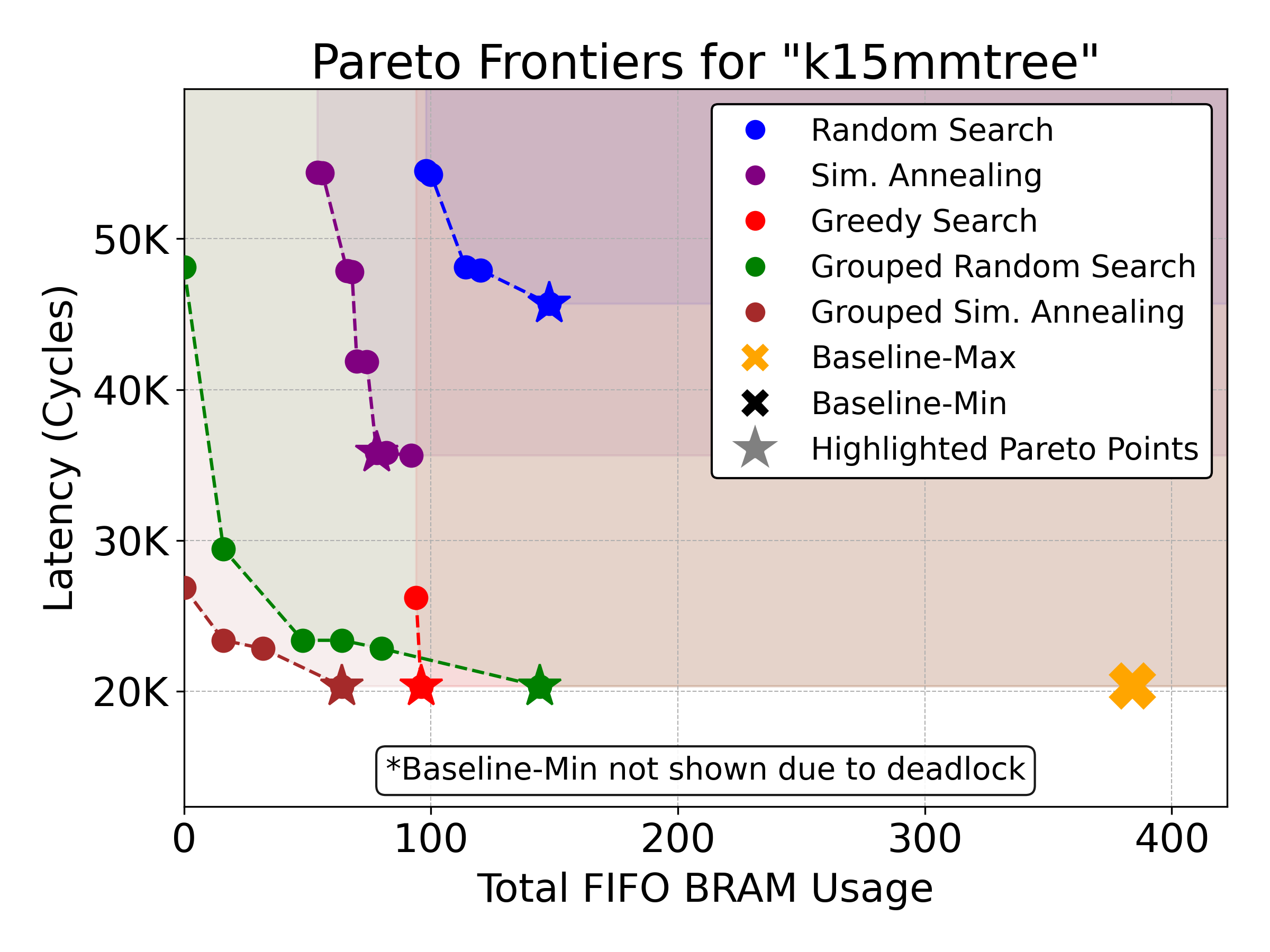}
    \includegraphics[width=0.30\linewidth]{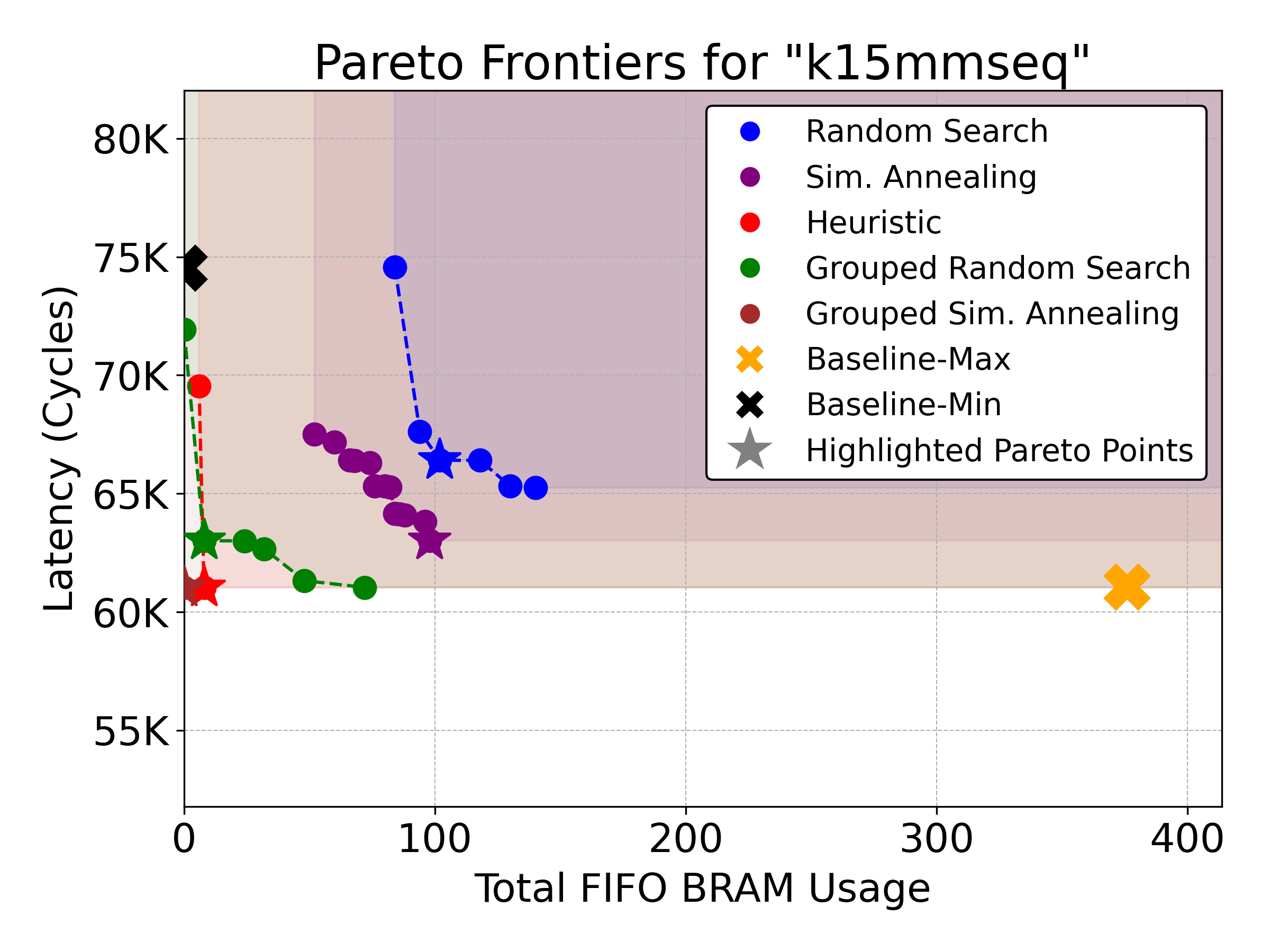}
    \includegraphics[width=0.30\linewidth]{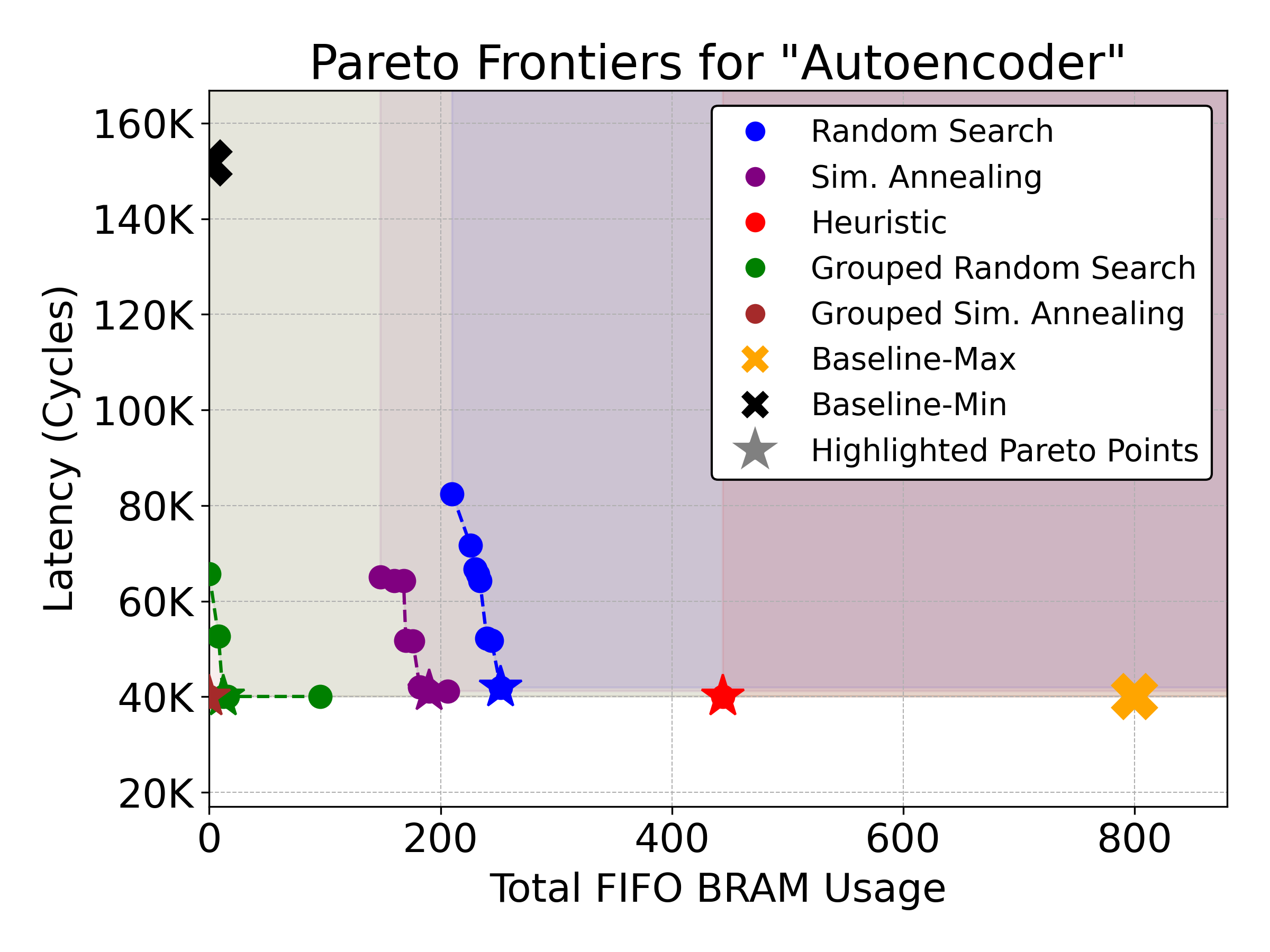}
    \vspace{-6pt}
    \caption{Pareto frontiers for the selected \textit{k15mmtree}, \textit{k15mmtseq}, and \textit{Autoencoder} designs. The ``highlighted Pareto points'' ($\bigstar$, using $\alpha = 0.7$) are used for comparison against the Baseline-Min (\protect\usym{2716}) and Baseline-Max points (\textcolor{baselinemax}{\protect\usym{2716}}) as described in \S\ref{sec:pareto}.}
    \label{fig:pareto}
\end{figure*}
\begin{figure*}[t!]
    \centering
    \vspace{-14pt}
    \includegraphics[width=0.90\linewidth]{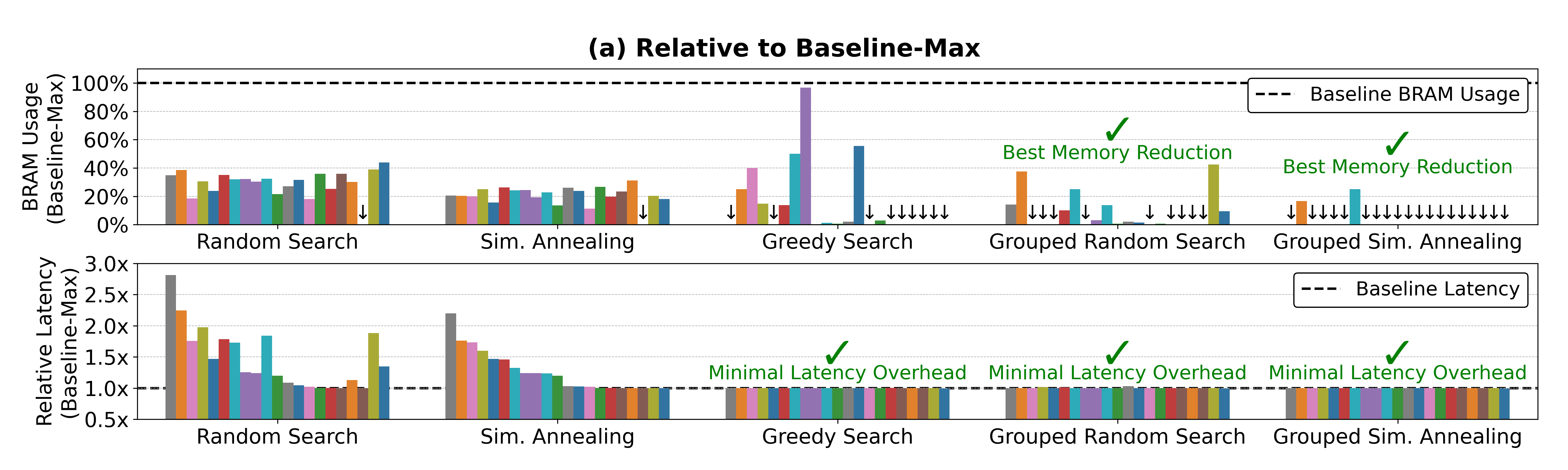}\\
    \vspace{-12pt}
    \includegraphics[width=0.90\linewidth]{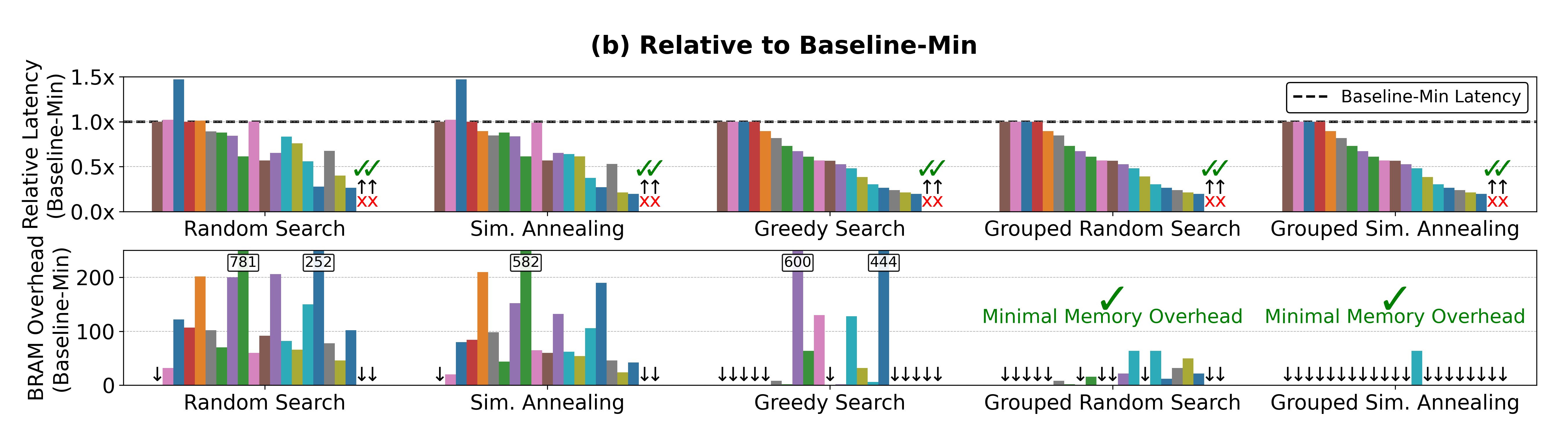}
    \vspace{-4pt}
    \caption{FIFO memory usage and relative latency of ``highlighted Pareto points'' ($\bigstar$)  relative to the na\"ively-sized baseline designs (\protect\usym{2716} and \textcolor{baselinemax}{\protect\usym{2716}}). Each bar represents a single design. For all metrics, lower is better. The down arrow \protect\contour{black}{$\downarrow$} indicates no BRAMs were used for all FIFOs in a given design. \textcolor{red}{\protect\usym{2715}}\,\protect\contour{black}{$\rightarrow$}\,\textcolor{Green}{\protect\usym{2713}} indicates that a Baseline-Min design was \textit{un}-deadlocked by \ourwork{}.}
    \label{fig:baseline}
    \vspace{-4pt}
\end{figure*}

\begin{table}
    \centering
    \footnotesize
    \renewcommand{\arraystretch}{0.80}
    \resizebox{0.47\textwidth}{!}{%
    \begin{tabular}{lcccc}
    \toprule
    Design & FIFOs & Co-Sim. & LightningSim & Diff \\
    \midrule
    atax & 175 & 2180 & 2181 & +0.0\% \\
    Autoencoder & 392 & 39178 & 40073 & +2.3\% \\
    bicg & 25 & 1112 & 1112 & \checkmark \\
    DepthSepConvBlock & 84 & 134541 & 134542 & +0.0\% \\
    FeedForward & 848 & 65997 & 65998 & +0.0\% \\
    gemm & 88 & 24051 & 24051 & \checkmark \\
    k2mm & 64 & 36352 & 36352 & \checkmark \\
    k3mm & 95 & 49092 & 49092 & \checkmark \\
    k7mmseq\_balanced & 112 & 5684 & 5684 & \checkmark \\
    k7mmseq\_unbal. & 108 & 10036 & 10036 & \checkmark \\
    k7mmtree\_unbal. & 128 & 8750 & 8750 & \checkmark \\
    mvt & 288 & 667 & 668 & +0.1\% \\
    ResidualBlock & 64 & 2092531 & 2092532 & +0.0\% \\
    k15mmseq\_imbal. & 59 & 7802 & 7802 & \checkmark \\
    k15mmseq & 188 & 61052 & 61052 & \checkmark \\
    k15mmseq\_relu\_imbal. & 116 & 8504 & 8504 & \checkmark \\
    k15mmseq\_relu & 232 & 28838 & 28838 & \checkmark \\
    k15mmtree\_imbal. & 163 & 16237 & 16237 & \checkmark \\
    k15mmtree & 192 & 20326 & 20326 & \checkmark \\
    k15mmtree\_relu\_imbal. & 340 & 16489 & 16490 & +0.0\% \\
    k15mmtree\_relu & 320 & 17277 & 17277 & \checkmark \\
    \bottomrule
    \end{tabular}%
    }
    \caption{The 21 benchmarks we used from Stream-HLS~\cite{streamhls}, including the number of FIFOs in each design, cycle counts from co-simulation and LightningSim, and their difference. A checkmark (\checkmark) indicates identical cycle counts.}
    \label{tab:cosim}
\end{table}

To evaluate our proposed technique, we need a collection of complex, realistic HLS designs that make heavy use of FIFO streams. The recent state-of-the-art work Stream-HLS~\cite{streamhls} presents a suitable case study for this purpose. We collect a set of 21 streaming designs open-sourced by the Stream-HLS authors, enumerated in Table~\ref{tab:cosim}, representing a wide variety of HLS kernels useful for machine learning.

We use AMD/Xilinx's Vitis HLS 2023.2 targeting the Alveo U280 FPGA platform to synthesize the designs. We then run them with LightningSim~\cite{lightningsim,lightningsimv2} version 0.2.6, generating an execution trace for each, which can then be used to generate a simulated cycle count for any FIFO depth. To check the accuracy of LightningSim and ensure its validity for our case study, we compare its cycle count output with the result of C/RTL co-simulation in Vitis HLS. In all but one of our designs, the LightningSim-computed latency is within \textbf{one cycle} of the true cycle count; one edge case (\texttt{Autoencoder}) is still within 2.3\% of the true cycle count.

For all results shown in Fig.~\ref{fig:pareto} and Fig.~\ref{fig:baseline}, each optimizer is given the same budget of 1,000 samples.

\subsection{Pareto Frontiers of Optimized Solutions}
\label{sec:pareto}

In Fig. \ref{fig:pareto}, we illustrate the Pareto frontiers generated by different optimizers relative to two baseline points: \textbf{Baseline-Max} and \textbf{Baseline-Min}. The Baseline-Max point represents a design where each FIFO is na\"ively maximally sized to fully buffer all the data that passes through it, which guarantees minimum latency. This is the default FIFO sizing that Stream-HLS adopts.
The Baseline-Min point represents a design with all FIFOs set to \textbf{depth=2}, the default smallest depth, guaranteeing minimum resource usage. By construction, the Baseline-Max solution is always deadlock-free, while the Baseline-Min solution can potentially deadlock. As shown in Fig.~\ref{fig:baseline} (indicated by \textcolor{red}{\protect\usym{2715}}), only two designs in our evaluation have a Baseline-Min solution that deadlocks, with \texttt{k15mmtree} as one of these kernels (also indicated in Fig.~\ref{fig:pareto}).

Qualitatively, we see that grouped optimizers, especially grouped simulated annealing, always find Pareto frontiers as good as or better than other optimizers and baselines. This is true for all designs we tested, including those not depicted.

\begin{table*}[th!]
\centering
\footnotesize
\renewcommand{\arraystretch}{0.1}
\resizebox{0.85\textwidth}{!}{
\begin{tabular}{l|c|ccccc}
\toprule
\multirow{2}{*}{\textbf{Design}} & \textbf{Co-Sim} & \multicolumn{5}{c}{\textbf{FIFO-Advisor}} \\
\cmidrule{2-7}
 & \textbf{(PAR=32)} & \textbf{Heuristic} & \textbf{Rnd.} & \textbf{Grp. Rnd.} & \textbf{SA} & \textbf{Grp. SA} \\
\midrule
atax & 0.61 days & 10.68 s. & 2.10 s. & 3.17 s. & 8.62 s. & 11.08 s. \\
Autoencoder & 8.68 days & 45.39 s. & 3.40 s. & 3.83 s. & 25.60 s. & 19.66 s. \\
bicg & 1.04 days & 0.00 s. & 0.10 s. & 0.28 s. & 0.10 s. & 0.13 s. \\
DepthwiseSeparableConvBlock & 2.06 days & 7.33 s. & 1.76 s. & 1.31 s. & 18.48 s. & 14.82 s. \\
FeedForward & 16.11 days & 74.20 s. & 3.94 s. & 5.17 s. & 26.01 s. & 25.58 s. \\
gemm & 7.05 days & 0.10 s. & 0.84 s. & 0.38 s. & 4.11 s. & 6.47 s. \\
gesummv & 0.46 days & 0.00 s. & 0.08 s. & 0.15 s. & 0.15 s. & 0.08 s. \\
k15mmseq & 1.82 days & 6.15 s. & 0.67 s. & 0.42 s. & 3.72 s. & 3.46 s. \\
k15mmseq\_imbalanced & 0.43 days & 0.55 s. & 0.29 s. & 0.28 s. & 1.52 s. & 1.34 s. \\
k15mmseq\_relu & 1.33 days & 11.90 s. & 0.78 s. & 0.82 s. & 10.29 s. & 7.15 s. \\
k15mmseq\_relu\_imbalanced & 0.37 days & 1.06 s. & 0.39 s. & 0.38 s. & 3.03 s. & 1.65 s. \\
k15mmtree & 1.59 days & 2.35 s. & 0.48 s. & 0.42 s. & 2.32 s. & 3.04 s. \\
k15mmtree\_imbalanced & 1.42 days & 1.98 s. & 0.41 s. & 0.49 s. & 2.56 s. & 4.09 s. \\
k15mmtree\_relu & 1.36 days & 13.37 s. & 1.00 s. & 0.81 s. & 7.99 s. & 7.54 s. \\
k15mmtree\_relu\_imbalanced & 1.33 days & 13.54 s. & 0.98 s. & 1.54 s. & 5.76 s. & 5.70 s. \\
k2mm & 8.46 days & 0.88 s. & 0.79 s. & 0.62 s. & 5.44 s. & 7.24 s. \\
k3mm & 11.79 days & 5.98 s. & 1.01 s. & 0.78 s. & 7.94 s. & 7.21 s. \\
k7mmseq\_balanced & 1.77 days & 1.11 s. & 0.60 s. & 0.36 s. & 1.31 s. & 1.28 s. \\
k7mmseq\_unbalanced & 2.59 days & 0.87 s. & 0.78 s. & 0.44 s. & 1.42 s. & 1.45 s. \\
k7mmtree\_balanced & 2.00 days & 0.47 s. & 0.46 s. & 0.31 s. & 0.93 s. & 1.18 s. \\
k7mmtree\_unbalanced & 2.23 days & 1.32 s. & 0.34 s. & 0.31 s. & 1.41 s. & 1.41 s. \\
mvt & 1.66 days & 0.13 s. & 1.44 s. & 0.68 s. & 7.68 s. & 10.81 s. \\
ResidualBlock & 9.41 days & 2.06 s. & 2.78 s. & 2.24 s. & 14.00 s. & 14.66 s. \\
ResMLP & 4.94 days & 141.41 s. & 5.77 s. & 4.65 s. & 26.22 s. & 45.87 s. \\
\midrule
\textbf{FIFO-Advisor Speedup Geomean} & & $\bm{10^{6.53}\pmb{\times}}$ & $\bm{10^{6.88}\pmb{\times}}$ & $\bm{10^{6.91}\pmb{\times}}$ & $\bm{10^{6.20}\pmb{\times}}$ & $\bm{10^{6.19}\pmb{\times}}$ \\
\bottomrule
\end{tabular}
}
\caption{Comparison of search runtime of \ourwork{} vs. traditional HLS/RTL co-simulation with a given budget of 1,000 samples; Note that, even with best-case estimates and allowing 32 parallel workers for co-simulation, \ourwork{} is multiple orders of magnitude faster; \textbf{Rnd}: Random Search, \textbf{Grp. Rnd}: Grouped Random Search, \textbf{SA}: Simulated Annealing, \textbf{Grp. SA}: Grouped Simulated Annealing, \textbf{PAR=32}: 32 parallel workers.}
\vspace{-2em}
\label{tab:cosim_fifo_opt_comparison}
\end{table*}

\subsection{Optimized Design Improvement}

The obvious question is, \textit{does \ourwork{} really improve HLS dataflow designs?} However, since \ourwork{} generates a frontier of Pareto-optimal FIFO configurations, there is no one way to answer this question---we must pick a design point from the Pareto frontier and compare it against a baseline configuration. To do this, we devise a simple scoring metric:
\begin{equation*}\alpha \left(\frac{\text{latency}}{\text{baseline latency}}\right)+(1-\alpha)\left(\frac{\text{FIFO BRAMs}}{\text{baseline FIFO BRAMs}}\right)\end{equation*}
Where $\alpha$ provides a customizable knob to trade off latency and memory. To select our ``optimal'' configuration for comparison, we choose $\alpha = 0.7$ relative to Baseline-Max, which prefers design points that better preserve the Baseline-Max latency even if they do not save as much memory. These selected points are highlighted as stars, $\bigstar$, in Fig. \ref{fig:pareto}, and will be used to compare against both baselines.

It is important to clarify that these comparisons to the selected ``optimal'' configuration based on $\alpha=0.7$ are for evaluation purposes only. \ourwork{} always solves for and provides the user with the full Pareto frontier of solutions, allowing the user to choose any configuration along this frontier based on their specific design and application trade-offs.

We first compare against Baseline-Max. Results are shown in Fig.~\ref{fig:baseline}(a). The random search and simulated annealing perform the worst: both optimizers frequently pick design points with substantial slowdowns (geomeans of 1.40\texttimes{} and 1.23\texttimes{} baseline latency, respectively) and less memory savings than other methods (average of 70.6\% and 79.4\% BRAM reduction). In contrast, our greedy search optimizer is able to preserve latency (0.9995\texttimes\footnote{Astute readers may notice that this geomean is less than 1\texttimes{} of Baseline-Max, which we claimed ``guarantees minimum latency.'' This is actually because by reducing FIFO sizes from Baseline-Max, some BRAM-backed FIFOs become shift registers instead, and shift register FIFOs have one fewer clock cycle of read delay than BRAM FIFOs, which can sometimes slightly decrease overall latency. This has no impact on the efficacy of \ourwork{} or our conclusions.}) and attain greater memory savings (85.6\%). The grouped optimizers identify many ideal configurations with zero BRAMs used by FIFOs and virtually no slowdown (1.0026\texttimes{} baseline latency for grouped random search, and 0.9994\texttimes{} for grouped simulated annealing). This supports the idea that arrays of streams behave similarly and can thus be optimized together.

We then compare against Baseline-Min; results are in Fig.~\ref{fig:baseline}(b). Foremost, we highlight that in all cases where Baseline-Min designs deadlock, \ourwork{} finds a \textit{non-deadlocked design with zero BRAM usage}; this alone is novel to \ourwork{}. Again, the random search and simulated annealing perform the worst: both optimizers can pick designs with some latency reduction (0.71\texttimes{} and 0.63\texttimes{}) but often with a high BRAM overhead (average of 131.0 and 97.7 BRAMs). In contrast, our greedy search optimizer is able to pick more designs with higher latency reduction (0.53\texttimes{}), but still incurs large BRAM overhead for some designs (67.4 BRAMs). As with before, the grouped optimizers identify many ideal configurations with significant latency reduction (0.53\texttimes{} and 0.52\texttimes{}) and virtually no BRAM overhead (13.9 and 3.0 BRAMs), with grouped simulated annealing as the best choice.

\subsection{Optimization Runtime}
\label{sec:results-runtime}

Across the board, we show that \ourwork{} has superior runtime performance solving the FIFO optimization problem compared to HLS/RTL co-simulation. To quantify this speedup, we estimate the co-simulation-based search runtime.

Running co-simulation for hundreds of FIFO configurations is prohibitively long, so we compute a conservative lower bound. We first run co-simulation once per design using maximum FIFO sizes (i.e. corresponding to Baseline-Max), which minimizes execution cycles due to reduced stalling. Since co-simulation runtime scales with cycle count, this configuration yields the fastest co-simulation runtime as our best-case estimate. We then multiply this best-case runtime by the number of configurations \ourwork{} explores for a given search technique to compute the total runtime for an equivalent co-simulation-based search. Results are also reported assuming perfect scaling with 32 parallel workers for co-simulation, with no overhead for task distribution or optimizer logic.

In contrast, we directly measure the runtime of \ourwork{}, which evaluates hundreds of FIFO configurations in seconds. As before, all optimizers are limited to exactly 1,000 samples, except for the heuristic approach, which deterministically chooses its own stopping point (ranging from ~10 to ~2,200 samples across designs in this evaluation).

Table \ref{tab:cosim_fifo_opt_comparison} summarizes runtime comparisons across all evaluated Stream-HLS designs, with speedups ranging from $\bm{10^5\pmb{\times}}$ to $\bm{10^7\pmb{\times}}$. Even under the most optimistic estimate, co-simulation-based search has no practical advantage. Critically, for designs with data-dependent control flow, only a simulation-based technique can solve the FIFO optimization problem with deadlock-free guarantees. With these runtime results, the choice between \ourwork{} and co-simulation is clear for all designs.

\begin{figure}[t]
    \centering
    \includegraphics[width=1.0\linewidth]{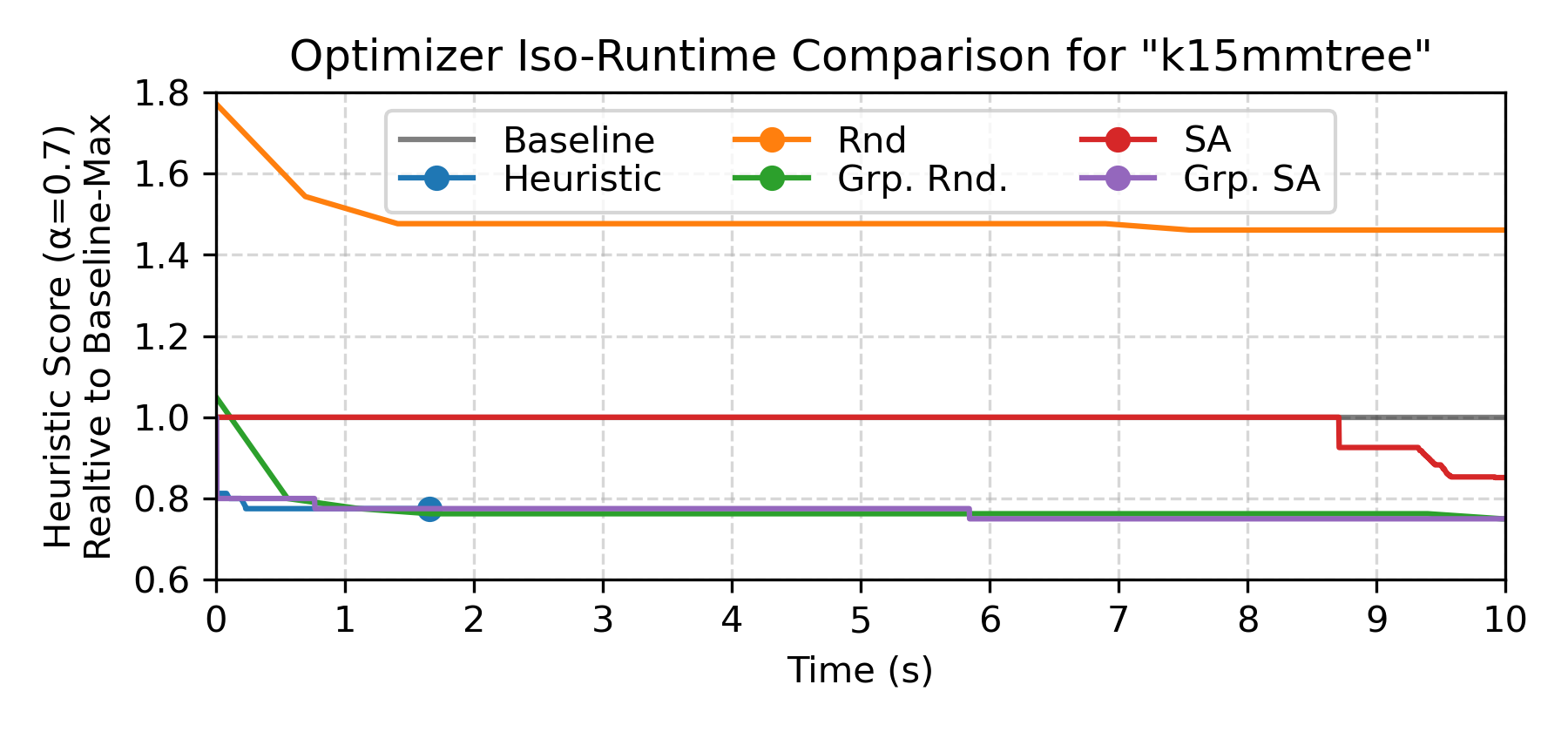}
\caption{Iso-runtime comparison of different optimizers on the \texttt{k15mmtree} design. Runtime includes optimizer logic overhead and calls to LightningSim to evaluate new FIFO configurations.}    \label{fig:runtime}
\end{figure}

We present additional search runtime results in Fig.~\ref{fig:runtime} to illustrate the convergence behavior of different optimizers on a selected design, \texttt{k15mmtree}, relative to the Baseline-Max FIFO configuration. Both grouped optimizers converge to the best solutions observed within 6 seconds, which we attribute to effective design space pruning. Notably, the heuristic optimizer also achieves near-optimal results within 2 seconds. These results demonstrate that \ourwork{} and the proposed optimizers consistently converge to high-quality solutions within seconds, making \ourwork{} ideal for designers who wish to iteratively prototype and optimize domain-specific HLS accelerators.

\subsection{Case Study: Data Dependent Control Flow}
\label{sec:cs-ddcf}

As discussed in \S \ref{sec:motivation-prior-work}, supporting FIFO optimization of designs with data-dependent control flow is crucial for supporting real-world hardware accelerator designs. Therefore, we demonstrate
\ourwork{}'s ability to optimize FIFOs in a complex hardware accelerator for graph neural networks from a prior work called FlowGNN~\cite{sarkar2023flowgnn}. These designs contain FIFOs for scattering and gathering data between nodes on a graph according to the graph connectivity passed in at runtime; this makes FlowGNN an ideal showcase for sizing many FIFOs with non-trivial deadlocking constraints and latency trade-offs.

Without loss of generality, we use FlowGNN's Principal Neighborhood Aggregation (PNA) design for our case study. We allocate a fixed budget of 5,000 samples per optimizer and all optimizer runs execute in less than 10s.

For this evaluation, the Baseline-Max configuration is modified to match the actual FIFO sizes used in the original FlowGNN PNA accelerator. These sizes were heuristically chosen by the HLS designer to prevent deadlock and minimize resource usage. Unlike Stream-HLS designs, the PNA FIFOs are not maximally sized, but the user-defined Baseline-Max configuration may still be suboptimal in balancing latency and resource utilization.

\begin{figure}[t]
    \centering
    \includegraphics[width=0.75\linewidth]{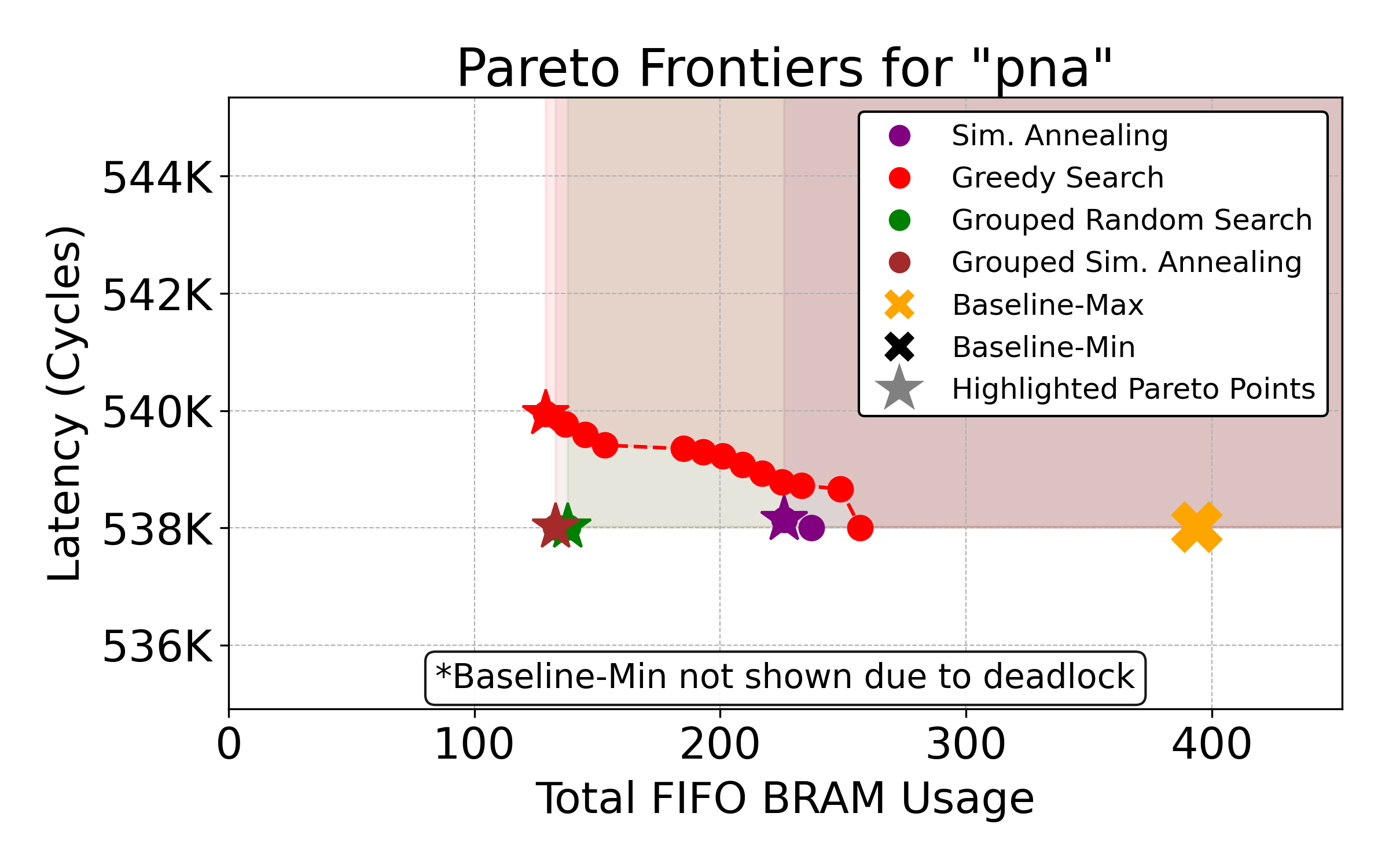}
    \caption{Pareto frontier for the selected \texttt{pna} design with data-dependent control flow. The ``highlighted Pareto points'' ($\bigstar$, using $\alpha = 0.7$) are used to compare against the Baseline-Max point (\textcolor{baselinemax}{\protect\usym{2716}}).}
    \label{fig:pna}
\end{figure}

Fig.~\ref{fig:pna} illustrates the Pareto frontiers generated by \ourwork{}'s different optimizers.  We can see that, like with our previous evaluations on designs with static control flow, our optimizers are able to find practical Pareto-optimal points relative to the user-sized Baseline-Max configuration.

A limitation of our current implementation is that we optimize FIFOs based only on one set of kernel inputs from the testbench; future work can easily extend our current approach by optimizing multiple executions jointly over a suite of test stimuli.

\section{Conclusion}

This work introduces \ourwork{}, the first DSE tool for rapidly discovering optimal FIFO sizes in HLS dataflow designs. Combining a fast, cycle-accurate HLS simulator, a FIFO memory usage model, and a multi-objective optimization formulation, \ourwork{} efficiently identifies a Pareto-optimal frontier of FIFO depth configurations for a given design, balancing latency and FIFO memory usage. We demonstrate that several of \ourwork{}’s optimizers can significantly reduce BRAM usage in real-world dataflow designs with minimal latency impact. We also show \ourwork{} can do so with minimal runtime and for designs with complex data-dependent control flow. Finally, we open-source our work as a standalone tool for HLS designers and integrate it with Stream-HLS to enable further multi-level optimization of dataflow designs.

\section*{Acknowledgments}
This work was partially supported by NSF awards CCF-1937599 and CCF-2211557, the CDSC industrial partners, the AMD\footnote{Jason Cong has financial interest in AMD} HACC Programs, the PRISM Center under the JUMP2.0 Program sponsored by Semiconductor Research Corporation (SRC) and DARPA, the Georgia Research Alliance based in Atlanta, Georgia under award GRA.25.039.GT.01.a, and NSF award CCF-2338365.

\bibliographystyle{IEEEtran}
\bibliography{refs}

\end{document}